\documentclass[preprint,preprintnumbers,amsmath,amssymb,superscriptaddress]{revtex4}

\usepackage{graphicx}
\usepackage{dcolumn}
\usepackage{amsmath}    
\usepackage{verbatim}   
\usepackage{color}      
\usepackage{subfigure}  
\usepackage{hyperref}   
\usepackage{bm}

\begin{document}


\title{Electric field induced coherent coupling of the exciton states in a single quantum dot
}

\author{A. J. Bennett}
\email{anthony.bennett@crl.toshiba.co.uk}
\affiliation{Toshiba Research Europe Limited, Cambridge Research Laboratory,\\
208 Science Park, Milton Road, Cambridge, CB4 OGZ, U. K.}

\author{M. Pooley}
\affiliation{Toshiba Research Europe Limited, Cambridge Research Laboratory,\\
208 Science Park, Milton Road, Cambridge, CB4 OGZ, U. K.}
\affiliation{Cavendish Laboratory, Cambridge University,\\
J. J. Thomson Avenue, Cambridge, CB3 0HE, U. K.}

\author{R. M. Stevenson}
\affiliation{Toshiba Research Europe Limited, Cambridge Research Laboratory,\\
208 Science Park, Milton Road, Cambridge, CB4 OGZ, U. K.}

\author{M. B. Ward}
\affiliation{Toshiba Research Europe Limited, Cambridge Research Laboratory,\\
208 Science Park, Milton Road, Cambridge, CB4 OGZ, U. K.}

\author{R. B. Patel}
\affiliation{Toshiba Research Europe Limited, Cambridge Research Laboratory,\\
208 Science Park, Milton Road, Cambridge, CB4 OGZ, U. K.}
\affiliation{Cavendish Laboratory, Cambridge University,\\
J. J. Thomson Avenue, Cambridge, CB3 0HE, U. K.}

\author{A. Boyer de la Giroday}
\affiliation{Toshiba Research Europe Limited, Cambridge Research Laboratory,\\
208 Science Park, Milton Road, Cambridge, CB4 OGZ, U. K.}
\affiliation{Cavendish Laboratory, Cambridge University,\\
J. J. Thomson Avenue, Cambridge, CB3 0HE, U. K.}

\author{N. Sk\"{o}ld}
\affiliation{Toshiba Research Europe Limited, Cambridge Research Laboratory,\\
208 Science Park, Milton Road, Cambridge, CB4 OGZ, U. K.}

\author{I. Farrer}
\affiliation{Cavendish Laboratory, Cambridge University,\\
J. J. Thomson Avenue, Cambridge, CB3 0HE, U. K.}

\author{C. A. Nicoll}
\affiliation{Cavendish Laboratory, Cambridge University,\\
J. J. Thomson Avenue, Cambridge, CB3 0HE, U. K.}

\author{D. A. Ritchie}
\affiliation{Cavendish Laboratory, Cambridge University,\\
J. J. Thomson Avenue, Cambridge, CB3 0HE, U. K.}

\author{A. J. Shields}
\affiliation{Toshiba Research Europe Limited, Cambridge Research Laboratory,\\
208 Science Park, Milton Road, Cambridge, CB4 OGZ, U. K.}

\date{\today}%



\maketitle 


\textbf{The signature of coherent coupling between two quantum states is an anti-crossing in their
energies as one is swept through the other. In single semiconductor quantum dots containing an
electron-hole pair the eigenstates form a two-level system which can be used to demonstrate quantum
effects in the solid state, but in all previous work these states were independent
\cite{Stevenson06, Jundt08, Muller09, Kowalik07, Geradot05}. Here we describe a technique to
control the energetic splitting of these states using a vertical electric field, facilitating the
observation of coherent coupling between them. Near the minimum splitting the eigenstates rotate in
the plane of the sample, being orientated at 45$^{o}$ when the splitting is smallest. Using this
system we show direct control over the exciton states in one quantum dot, leading to the generation
of entangled photon pairs.}

It is well known that the exchange interaction in single semiconductor dots results in the exciton
eigenstates being linearly polarised with an energy difference known as the fine-structure
splitting (FSS, $|s|$) \cite{Bayer99, Gammon96}. The magnitude of the FSS is determined by
anisotropy in the strain, shape and composition of the dot, in addition to a contribution from the
crystal inversion asymmetry \cite{He08}. When the FSS is smaller than the linewidth, the
biexciton-to-exciton-to-empty cascade can lead to the emission of polarisation-entangled
photon-pairs \cite{StevensonNature06, Hafenbrak08}, which has motivated much work on this subject.

Recently, theoretical work has suggested that for realistic strain-tuned dots a minimum in the FSS
($s_{0}$) will be observed of the order of 3 $\mu$eV, due to the symmetry of the crystal
\cite{Singh10}, but this has yet to be confirmed by experiment. Some success at tuning the FSS has
been reported using strain \cite{Seidl06} but this did not reach zero. Other tuning techniques such
as magnetic field \cite{Bayer99, Stevenson06}, strong coherent lasers \cite{Jundt08, Muller09},
lateral electric field \cite{Kowalik07, Geradot05, Vogel07} and vertical electric field
\cite{Kowalik06, Marcet10} have been investigated but those that have been able to minimise the FSS
have seen the states cross \cite{Stevenson06, Jundt08, Muller09, Kowalik07, Geradot05}.

\begin{figure}[h]
\includegraphics[width = 70mm]{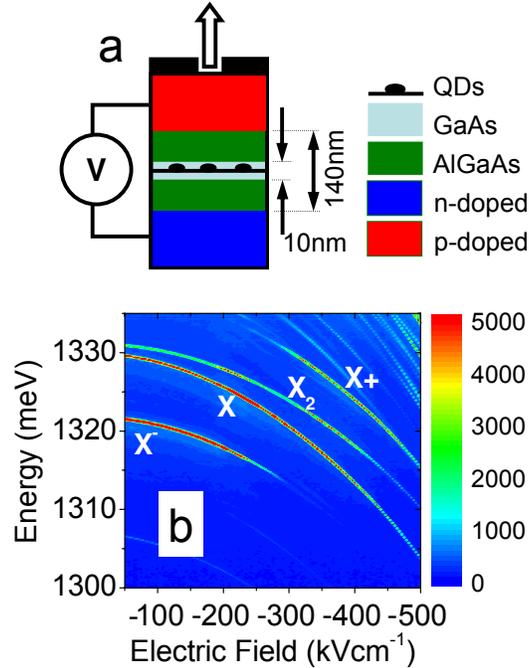} 
\caption{\label{Fig1} Device design and observed giant stark-shift of the excitonic transitions.
\textbf{a} heterostructure design. \textbf{b} typical plot of photoluminescence versus electric
field for a single quantum dot.}
\end{figure}

Arguably tuning the FSS with vertical electric field is the most practical technique yet reported,
but the low confinement energies have limited the fields that can be applied to a few tens of
kVcm$^{-1}$ before carriers tunnel from the dot, so relatively small changes in FSS were observed
\cite{Kowalik06, Marcet10}. Here we demonstrate a design of heterostructure (Figure \ref{Fig1}a)
that allows much larger electric fields to be applied. Then both eigenstates of the exciton Stark
shift at different rates, leading to a linear change in the FSS of over 100 $\mu$eV. At appropriate
fields we observed for the first time anti-crossings in the energies of the two exciton levels.
Near the anti-crossing hybridisation of the polarisation states of the upper and lower branches
leads to their rotation in the plane of the sample. In a dot that has a small anti-crossing (below
the homogeneous linewidths) we are thus able to demonstrate control of entangled photon pair
emission using electric field. Other dots exhibit a large minimum splitting (FSS greater than the
homogeneous linewidths) and a coherent superposition of the usual exciton states is observed.

\begin{figure}[h]
\includegraphics[width=80mm]{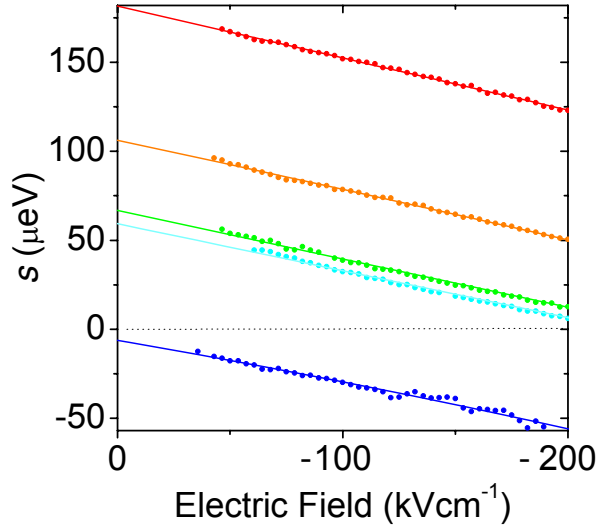}
\caption{\label{Fig2} Fine-structure splitting $s$ as a function of electric field for 5 dots with
naturally different $s$ at low field, showing $s$ varies in the same manner for all dots in this
sample. One dot has an inverted fine-structure at all fields, with the lowest energy exciton state
orthogonal to other dots we have studied, and is plotted with a negative $s$.}
\end{figure}

As electric field is applied all transitions Stark shift with the applied field (Fig. \ref{Fig1}b),
following the form $E$ = $E_{0}+ p F +\beta F^{2}$, where $p$ is the permanent dipole moment in the
z-direction, $\beta$ the polarisability and $F$ the applied field \cite{Barker00, finley04,
Vogel07}. At large values of FSS the exciton eigenstates can be thought of as radiating dipoles
aligned along the $[110]$ and $[1\bar{1}0]$ crystal axes (see supplementary information), whose
orientation is mapped onto the emitted photon's polarisation ($X_{H/V}$ denoting photons of
horizontal or vertical polarisation). Remarkably, away from zero FSS all neutral states display a
linear change in the magnitude of the FSS with electric field that has gradient $\gamma$ = -0.285
$\pm$ 0.019 $\mu$eV kV$^{-1}$cm (Figure \ref{Fig2}). Even the unusual dot we have identified that
has its lowest energy exciton eigenstate orthogonal to all others in this sample, and thus is
plotted on Fig. 2 with a negative $s$, has the same gradient. This value of $\gamma$ is independent
of the FSS at zero field, emission energy, binding energy and the Stark shift parameters $p$ and
$\beta$. The observed linear shift may be explained by the fact that the two eigenstates of the
neutral exciton have slightly different confinement potentials in the two directions. This leads to
different permanent dipole moments along the z-direction \cite{finley04} giving $p_{H} - p_{V} =
\gamma$. However, the polarisability of these states is unaffected by this in-plane anisotropy
\cite{Barker00} and is controlled only by the height of the confinement potential, which is the
same for both $X_{H}$ and $X_{V}$. Thus the measured value of $\gamma$ means the dipole moments of
the two exciton eigenstates must differ by a few percent. The fact that $\gamma$ is so similar in
our ensemble suggests the dots all have comparable in-plane anisotropy, and that this value could
be manipulated by changing the shape of the dots.

\begin{figure}[h]
\includegraphics[width=150mm]{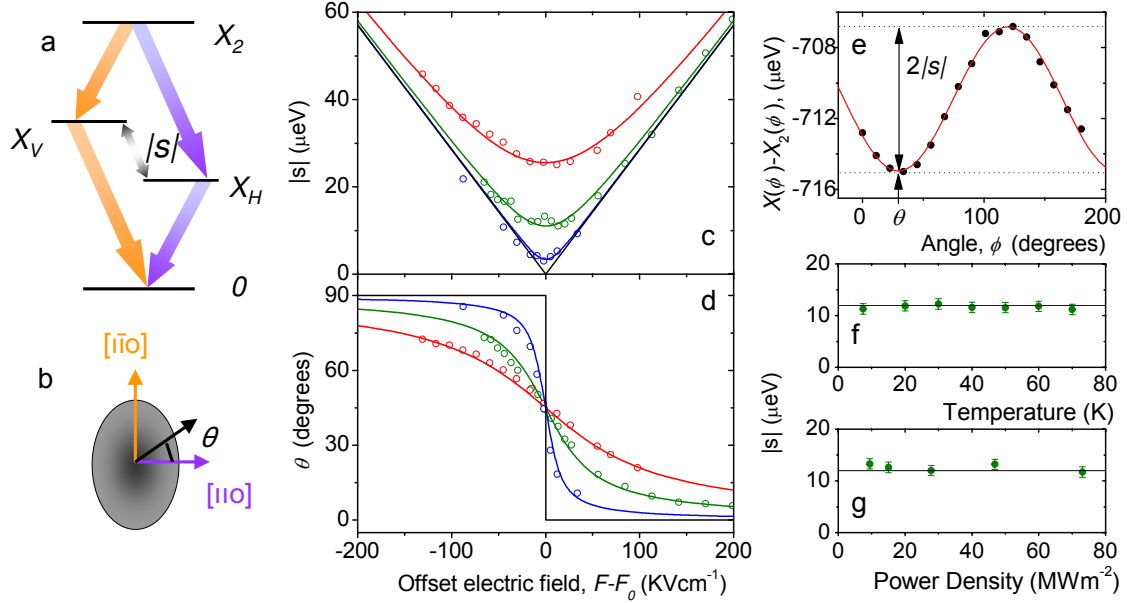} \caption{\label{Fig3} Characteristics of electric-field-induced coherent coupling
 of exciton states in a single quantum dot. \textbf{a} Energy levels in the neutral cascade of
 a single dot. \textbf{b} Orientation of the eigenstates relative to the crystal axes. \textbf{c} and \textbf{d} plot $|s|$ and $\theta$
as a function of the electric field (offset by $F_{0}$)
  for three quantum dots with values of $s_{0}$ equal to 25.5, 12.0 and 3.0 $\mu$eV and $F_{0}$ of -140.5, -234.5 and -98.0 kVcm$^{-1}$. The black line shows what would be
  expected from a dot with vanishing magnitude of anti-crossing. \textbf{e} The result of a typical measurement of the $[X(\phi)-X_{2}(\phi)]$
 from which we can extract $|s|$ and $\theta$. \textbf{f} and \textbf{g} measurements of the amplitude of the anti-crossing for one of the dots
 as a function of temperature and excitation intensity, respectively.}
\end{figure}

This ability to continuously tune the FSS over such a large range allows us to observe an
anti-crossing in the neutral excitonic levels. The variation of the FSS with field is shown in
Figure 3c for three dots, clearly indicating coherent coupling between these states. The splitting
is well described by a simple model:
\\
\linebreak
\begin{subequations}
$E\left(
    \begin{array}{c}
      cos\theta \\
      sin\theta \\
    \end{array}
  \right)
=\left(
   \begin{array}{cc}
     E_{0} & s_{0}/2 \\
     s_{0}/2 & E_{0}-\gamma (F-F_{0}) \\
   \end{array}
 \right)
 \left(
    \begin{array}{c}
      cos\theta \\
      sin\theta \\
    \end{array}
  \right)$
\end{subequations}
\\

where $s_{0}/2$ quantifies the coupling between the states, $F$ is electric field, $F_{0}$ the
field at minimal splitting and the energy levels move together with a rate $\gamma$ in the absence
of the coupling term $s_{0}/2$. This is the same form of equation that is used to describe coupled
harmonic-oscillators \cite{Feynman}, strong light-matter coupling \cite{Khitrova06}, and
anti-crossings in the states of molecular systems \cite{Stinaff06}. When $\gamma(F-F_{0})\ggg
s_{0}$ the natural basis to choose is that aligned with the crystal axes, and nearer $(F-F_{0})=0$
the eigenstates will be a coherent mixture, with components $sin\theta$ and $cos\theta$. In this
system the parameter $\theta$ is a real angle that describes the orientation of the eigenstates
relative to the crystal lattice (Figure \ref{Fig3}b). The eigenvalues, $E_{\pm}$, and eigenvectors
are well known, with
\\
\linebreak
\begin{subequations}
$E_{\pm}=E_{0}-\frac{\gamma(F-F_{0})}{2} \pm \frac{1}{2}\sqrt{\gamma^{2}(F-F_{0})^{2}+s_{0}^{2}}$
\\
 $\theta = \pm tan^{-1}\left[\frac{s_{0}}{\gamma (F-F_{0}) \pm (E_{-}-E_{+})}\right]$
\end{subequations}
\\

We note that this simple model has degenerate solutions where the eigenstates rotate either
clockwise or anti-clockwise when they approach $(F-F_{0})=0$. However, in practice we observe each
dot has a clear preference to rotate one way or another in the plane of the sample, but never into
the circular basis. Our measurements are consistent with the ensemble having no preferred direction
of rotation. The origin of this handedness in individual dots is unknown, but may be determined by
local defects or fields in the semiconductor.

Experimental data for three dots with varying sizes of anti-crossing and $F_{0}$ are shown in
Figure \ref{Fig3}c and d, where all three dots rotate in the same direction. This data shows
excellent agreement between the simple model and the experiment, indicative of a coherent coupling
that can be activated with electric field.

A study of 22 dots revealed values of $s_{0}$ in the range 0.7 to 42.9 $\mu$eV, with lower values
being observed more frequently (see supplementary information). No trend was observed between the
magnitude of this value and other parameters associated with the electronic states of the dot, such
as the FSS at $F = 0$, $F_{0}$, the emission energy or $(X(\phi)-X_{2}(\phi))$.

Coupled pairs of oscillators have two normal mode frequencies whose values depend in part on the
damping, or decoherence, affecting those individual oscillators \cite{Feynman}. For this reason we
have investigated varying the sample temperature (from 4K to 70K, Figure \ref{Fig3}f) and the
excitation power, changing the ratio of exciton to biexciton intensity from 10:1 to 1.3:1, (Figure
\ref{Fig3}g) for the dot with $|s_{0}|$=12.0 $\mu$eV. Both of these factors will vary the
decoherence experienced by a single state \cite{Berthelot06}, but have no effect on the magnitude
of the coupling. This suggests that although these external factors may cause decoherence of the
individual states (such as the $T_{2}$ time) it does not affect the time-scale on which the
two-states dephase relative to each other \cite{Hudson07}.

\begin{figure}[h]
\includegraphics[width=120mm]{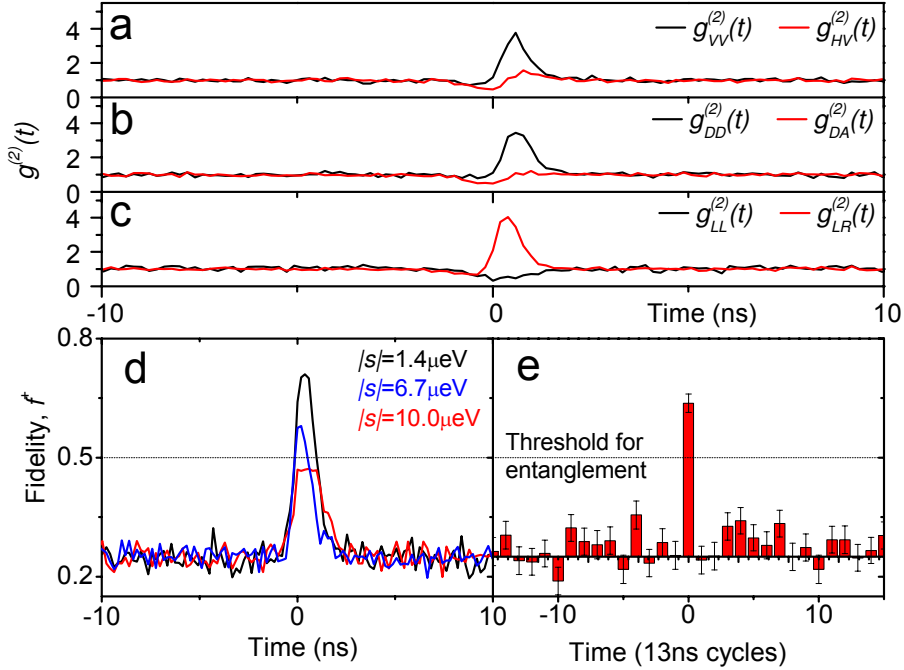}
\caption{\label{Fig4} Entanglement in the cascade emission of a dot with $s_{0}$ = 1.5 $\mu$eV.
Polarised cross-correlations between $X$ and $X_{2}$ photons in three orthogonal polarisation
bases: \textbf{a} rectilinear $\{H,V\}$, \textbf{b} diagonal $\{D,A\}$ and \textbf{c} circular
$\{L,R\}$. \textbf{d} Fidelity of the emitted state, for three different values of $|s|$.
\textbf{e} Fidelity of the emission with pulsed optical excitation, considering only those X
photons emitted within 100ps of a preceeding biexciton photon.}
\end{figure}

We now show that we are able to generate entangled photons from a dot with a FSS of over 50 $\mu$eV
at zero field, by simply applying a voltage to the sample. The dot we study has an anti-crossing of
amplitude $s_{0}$ = 1.5 $\mu$eV at a field of -240 kVcm$^{-1}$. This splitting is below that
required to observe entangled photon emission from the cascade \cite{StevensonNature06,
Hafenbrak08, Hudson07}. To confirm this, polarised cross-correlation measurements were made between
the $X$ and $X_{2}$ transitions at $|s|=s_{0}$ (Figure \ref{Fig4}a-c). The shapes of the peaks are
dominated by the gaussian-like instrument response function with width $\sim$600 ps. As expected,
strong correlation is observed in the rectilinear and diagonal bases (Figure \ref{Fig4}a and b),
and strong anti-correlation in the circular basis (Figure \ref{Fig4}c), when the two photons are
emitted at closely spaced times. The fidelity of the emission to the Bell state $\Psi^{+} =
[|X^{H}X_{2}^{H}\rangle+|X^{V}X_{2}^{V}\rangle]/\sqrt{2}$, $f^{+}$, is given by
$[C_{\{H,V\}}+C_{\{D,A\}}-C_{\{L,R\}}+1]/4$ \cite{Hudson07} where $C_{\{H,V\}}$ denotes the degree
of polarisation correlation in the $\{H,V\}$ basis. We obtain $f^{+}$ = 71 $\pm$ 3$\%$. Similar
measurements with pulsed excitation give $f^{+}$=64 $\pm$ 3$\%$ when averaging over all photons
emitted within 100 ps of each other (Figure \ref{Fig4}e). At this finite splitting the basis state
that has maximum $f^{+}$ rotates at a rate proportional to $|s_{0}|$, which combined with
re-excitation causes $f^{+}$ to fall to the classical value of 0.25 at times away from zero
\cite{Stevenson08}. In both pulsed and CW measurements at $s_{0}$, $f^{+}$ is above the threshold
of 0.5 confirming the emission of entangled photon pairs has not been degraded by the application
of such a large electric field. At different values of electric field and increased $|s|$ CW
correlation measurements confirm that as spectral distinguishability is introduced into the cascade
$f^{+}$ falls as expected (Figure \ref{Fig4}d).

We have demonstrated an effective and versatile technique to control the FSS of a single quantum
dot, which has enabled us to observe anti-crossings in the fine-structure of single dots. Such
electric-field control is well suited to the incorporation of high-quality, low volume cavities
facilitating higher efficiencies and cavity QED effects \cite{Ellis07, Bockler08}. In future, this
technique will allow control of the fine-structure splitting and eigenstates of the exciton on a
time scale faster than the radiative lifetime \cite{Bennett05, McFarlane10}, enabling manipulation
of superpositions stored in these states.

\section{Methods}

The InAs dots are grown at the centre of a 10nm GaAs quantum well clad with a short period
superlattice equivalent to Al$_{0.75}$Ga$_{0.25}$As. Doping that extends into the superlattice
allows application of on electric field along the growth direction. This $p-i-n$ device has a $d$ =
140nm thick $i$-region and is encased in a weak planar microcavity consisting of 14/4 periods
below/above the dot-layer. Thus the applied electric field is calculated to be $(V-V_{bi})/d$ where
the built-in potential, $V_{bi}$, is 2.2V. The Al$_{0.75}$Ga$_{0.25}$As barrier on either side of
the quantum well controls the charging of the dot. Although the dot layer is positioned an equal
distance from the n and p contact the tunneling rates will differ substantially due to the lower
effective mass of the electron and different confinement energies of the carriers.

Excitation and photon collection occurs through an opaque metallic film on the sample surface,
patterned with micron-diameter apertures (Figure \ref{Fig1}a). During spectroscopy the samples were
excited by a CW 850 nm laser diode, which creates carriers in the wetting layer and dot only. For
cross-correlation measurements a Ti-Sapphire laser operating in either pulsed or CW regime excited
the dots at 850 nm.

Single dots have a characteristic spectral arrangement of optical transitions that can readily be
identified as exciton ($X$), biexciton ($X_{2}$) and charged excitons ($X^{+}$,$X^{-}$), where this
notation refers to the initial state. For neutral $X_{2}$-to-$X$-to-empty cascade we are able to
induce Stark shifts of 25 meV at 500 kVcm$^{-1}$. Study of several dozen dots emitting in the range
1310-1340 meV showed the FSS at 50 kVcm$^{-1}$ displayed a Gaussian distribution centred on 109
$\mu$eV with width 67 $\mu$eV. Such a distribution fits well with the trend previously reported for
dots in GaAs\cite{Young05}. These observations confirm that placing the dots in a quantum well has
not changed their electronic properties.

We determine $|s|=|E_{+}-E_{-}|$ and $\theta$ by measuring spectra polarised at multiple angles
$\phi$ to the $[110]$ crystal direction, extracting the energy difference between the exciton and
biexciton transitions, $(X(\phi)-X_{2}(\phi))$. This technique eliminates small energy shifts
induced by rotation of the polarisation optics giving measurements of $s(\phi)$ with sub-$\mu$eV
accuracy \cite{Young05, Stevenson06}. When $|s|$ is below the resolution of the system used for
this measurement ($\thicksim$40 $\mu$eV) we observe a sinusoidal variation in
$(X(\phi)-X_{2}(\phi))$ (Figure \ref{Fig3}e) from which we determine the orientation angle,
$\theta$ and magnitude of the FSS, $|s|$, at each field.

\section{Acknowledgements}

This work was partly supported by the EU through the Integrated Project QESSENSE and the Marie
Curie Framework 7 project Spin-Optronics. EPSRC and TREL provided support for RBP, MP and ABDLG and
QIPIRC for CAN. We would like to thank D. Granados and R. Young for useful discussions on design of
the device.

\section{Author Contributions}
The samples were grown by I.F, C.A.N, D.A.R and processed by R.B.P. The optical measurements were
made by A.J.B, M.P and R.M.S. A.J.S. guided the work. All authors discussed the results and their
interpretation. A.J.B wrote the manuscript, with contributions from the other authors.


\end{document}